# Efficient asymmetric transmission of elastic waves in thin plates with lossless metasurfaces

Bing Li[1], Yabin Hu[1], Jianlin Chen[3], Guangyuan Su[2], Yongquan Liu[2, *], Meiying Zhao[1], Zheng Li[3]

[1] *School of Aeronautics, Northwestern Polytechnical University, Xi'an, Shaanxi, 710072, China*

[2] *State Key Laboratory for Strength and Vibration of Mechanical Structure, School of Aerospace Engineering, Xi'an Jiaotong University, Xi'an, 710049, China*

[3] *Department of Mechanics and Engineering Science, Peking University, Beijing 100871, China*

*Corresponding author E-mail address: liuy2018@xjtu.edu.cn*

**Abstract**

Requiring neither active components nor complex designs, we propose and experimentally demonstrate a generic framework for undistorted asymmetric elastic-wave transmission in a thin plate just using a layer of lossless metasurface. The asymmetric transmission stems from the uneven diffraction of +1 and -1 orders on opposite sides of the metasurface, respectively. Compared with previous loss-induced strategies, the present metasurface maintains a nearly total transmission for the transportation side, but a total reflection from the opposite side, exhibiting a higher contrast ratio of transmission. Moreover, we illustrate that this strong asymmetric behavior is robust to the frequency, the incident angle and the loss effect. The present work paves new avenues to compact rectification, high resolution ultrasonography, vibration and noise control in elastodynamics and acoustics.

Nonreciprocal wave transmission is fundamental to various rectifying scenarios of the directional-dependent energy flux [1-5]. Analogous to the electromagnetic/optical field, asymmetric acoustic/elastic-wave transmission has recently been realized based on bulk artificial materials under the effect of nonlinearity [5, 6], active components [7, 8], structural symmetry-breaking [9] and topological insulators [10,11]. However, the problem of either inherent signal distortions, nontrivial biasing forces, large-volume structures, or narrow operating band still poses challenges for a further practical implementation. Moreover, existing designs mostly show weak unidirectional effects due to insufficient transmission amplitude on the propagating side, which is a crucial issue and has not been effectively addressed so far [12].

Meanwhile, the metasurface [13-28], considered as a 2D counterpart of metamaterials from the function point of view, has opened up new possibilities for realizing extraordinary wave manipulation in the form of lightweight and thin planar designs. Specifically, for the asymmetric wave control, a design combining double-layer metasurface structures has been proposed for asymmetric acoustic transmission [29]. More recently, a pioneering work has been reported on the asymmetric acoustic transmission by utilizing single-layer, lossy passive metasurfaces based on loss-induced suppression of high order diffraction [30]. However, the lossy effect leads to an obvious attenuation on the transmitted sound waves from both incident sides [30, 31]. How to realize undistorted "one-way" performance using a compact passive structure, that also offers high transmission of the required signal simultaneously, is still an open challenge. This challenge seems more striking in elastodynamics due to additional wave modes and the inherent mode conversion of elastic waves.

Here, inspired by the integer parity feature of acoustic metasurface [27], we develop a generic framework and experimental realization of undistorted asymmetric transmission of guided waves in a thin plate using a layer of lossless elastic metasurfaces. Requiring neither active control systems nor complex microstructures, our approach allows us to demonstrate a strong asymmetric performance, maintaining a nearly total transmission for one-side incidence, but total reflection for another side. More remarkably, the asymmetric effect is loss-independent, broadband and valid within a wide range of incident angles.

Figure 1 shows the schematic diagram illustrating the control strategy of asymmetric elastic-wave transmission. A passive, lossless elastic metasurface composed of periodically repeated phase gradient supercells is considered, as depicted in Figs. 1(a) and 1(b). For each supercell, it includes $m$ unit cells, across which the transmitted phase shift can realize a full coverage of $2\pi$ per period. The directions of outgoing (reflected or transmitted) waves are determined by [17, 27]

$$\sin\theta_{\text{in}} + n\frac{\lambda}{L} = \sin\theta_{\text{out}} \tag{1}$$

where $\theta_{\text{in}}/\theta_{\text{out}}$ is the angle of incident/outgoing waves, $\lambda$ is the wavelength, $L$ is the length of the supercell, and $n$ is an integer representing the diffraction orders. For the case of $n = 1$, it is the well-known generalized Snell's law (GSL) established by Yu et al. in 2011 [13]. If we set $L \leq \lambda$, only three diffraction orders ($n = \pm 1$ and 0) of the scattering waves can be obtained under different incident angles (see Fig. 1(c)). Specifically, it is readily to select two opposite incident angles of $\pm\theta_{\text{in}}$ to fulfill that for the negative incidence (NI), $-\theta_{\text{in}}$ is smaller than the critical angle of $\theta_c = \sin^{-1}(1 - \lambda/L)$, while for the positive incidence (PI), $+\theta_{\text{in}}$ is larger than $|\theta_{in}|$. In consideration of the small amplitude of the $0^{th}$ order wave, the dominant propagation modes for NI and PI will be of the $+1$ and $-1$ orders, respectively, as illustrated in Fig. 1(c). Starting from the two kinds of diffraction orders, we will control the asymmetric elastic-wave transmission in lossless metasurfaces.

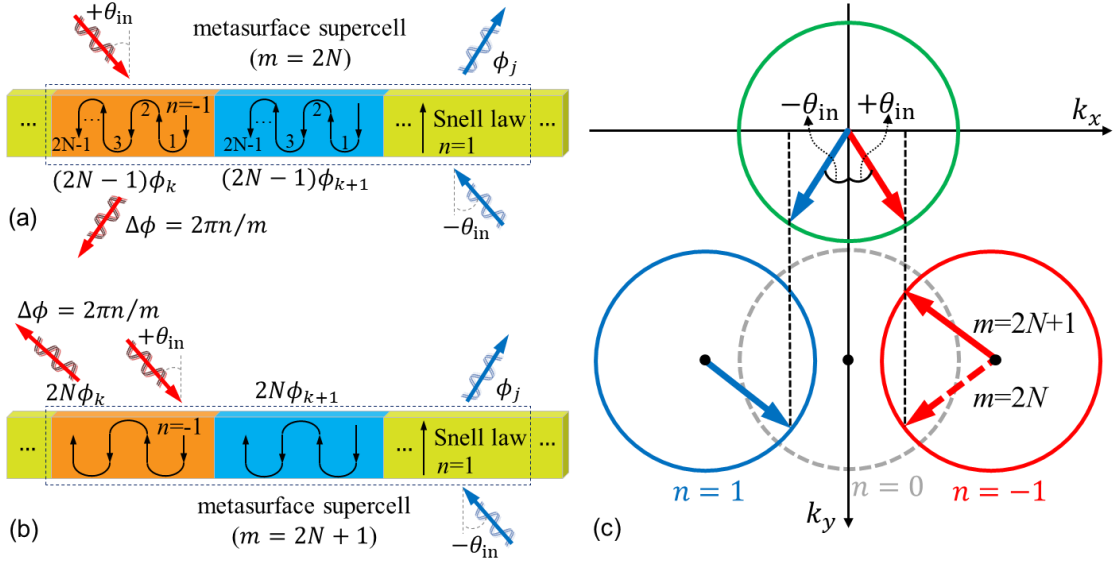

Fig. 1. Schematic of wave propagation through a lossless elastic metasurface consisting of periodically repeated phase gradient supercells (a) with an even number of $2N$ unit cells and (b) with an odd number of $2N+1$ unit cells per period. (c) Equifrequency contours of the elastic metasurface. The green circle is the equifrequency contour of incident wave. The red, gray and blue circles correspond to the contours of the -1, 0 and 1 order, respectively.

Based on GSL for NI, the dominant scattering waves of the +1 order will pass directly through the metasurface across the $j^{th}$ unit cell with transmitted phase shift $\phi_j = \frac{2\pi j}{m}$ (see Figs. 1(a) and 1(b)). Nevertheless, for PI, the dominant waves of the $-1$ order will obey the mechanism of multiple reflections, leading to two possibilities: 1) When the number of multiple reflections is odd, i.e. $2N-1$, the scattering waves will escape from the transmitted interface after $2N-1$ times reflection (see Fig. 1(a)). The multiple reflection mechanism, together with the utilization of lossy materials, has been proposed to design asymmetric acoustic metasurface by Li et. al [30]. However, for lossless metasurfaces, this multiple reflection may only give rise to a phase delay in transmission, but the transmitted fields from the two opposite incident directions are still symmetrical. 2) It is more remarkable that when the metasurface provides an even number of $2N$ reflections, the scattering waves will propagate back to the incident side (reflected interface) (see Fig. 1(b)), and almost no transmission occurs, which is totally different from the transmission of NI. Correspondingly, asymmetric elastic-wave transmission can be achieved at this pair opposite incident angles $\pm\theta_{in}$ by even a lossless metasurface. Although no requirement of lossy materials in this method, it should be pointed out that most lossy metasurfaces also work here due to the following reasons: on one hand, the use of ultrathin metasurface makes a high transmission for NI unless the loss is extremely large; on the other hand, the multiple reflection of PI will reduce the transmission of the unwanted $0^{th}$ order mode, which further benefits the asymmetric wave propagation. The effects of loss will be investigated in detail later.

Based on this strategy, it is evident that the integer parity control of multiple reflections plays a crucial role in asymmetric elastic-wave transmission, which can be realized by the supercell design. As illustrated in Fig. 1(a), after odd $(2N - 1)$ times of reflection, the phase shift in two adjacent unit cells can be expressed as $(2N - 1)\phi_k$ and $(2N - 1)\phi_{k+1}$, respectively, where $\phi_k$ and $\phi_{k+1}$ are the phase shift in the two adjacent unit cells after one-time reflection. In view of the supercell consists of $m$ unit cells, the phase difference between adjacent unit cells is $\Delta\phi = (2N - 1)(\phi_{k+1} - \phi_k) = (2N - 1)\frac{2\pi}{m}$. Meanwhile, for the $n^{th}$ order scattering waves, the phase difference of two adjacent unit cells can be written as $\Delta\phi = \frac{2\pi n}{m}$. Considering that the negative order of diffraction $(n = -1)$, a phase wrap of $2\pi$ is introduced to build the self-consistent couple of the two identical phase differences, i.e., $\frac{-2\pi}{m} + 2\pi = (2N - 1)\left(\frac{2\pi}{m}\right)$, leading to $m = 2N$. Similarly, for even times of multiple reflections (see Fig. 1(b)), we can obtain $\frac{-2\pi}{m} + 2\pi = 2N\left(\frac{2\pi}{m}\right)$, leading to $m = 2N + 1$. In short, we can control multiple reflections by changing the number of the unit cells in each supercell, which can further control the asymmetric phenomenon. Peculiarly, asymmetric elastic-wave transmission can be realized in a lossless metasurface when an odd number of unit cells are used to shape a supercell, whereas the transmission still keep symmetric in the lossless metasurface with an even number of unit cells per period.

A series of numerical simulations and experimental studies are performed to verify this idea. We begin with a representatively lossless metasurfaces consisting of periodically arranged supercells including an odd number ($m = 3$) of zigzag-like unit cells per period, as depicted in the inset of Fig. 2(a). The width of the elastic metasurface $l$ equals $0.67\lambda$ at the operating frequency of 15 kHz ($\lambda = 30.6$ mm). The phase shift and amplitude of transmitted waves across the proposed unit cells are provided in Supplementary Material [32]. Figure 2(a) shows the calculated out-of-plane displacement field contours in a thin steel plate for oblique incidences at $\pm 30°$. In the positive/negative incidence direction, the elastic-waves are almost totally reflected/transmitted. Thus the stark difference between the transmission properties for the two oppsite directions is clearly exhibited, which confirms that one can achieve efficient asymmetric transmission just using a lossless metasurface. For comparison, as illustrated in Fig. 2(b), waves in both directions are almost totally transmitted in the lossless metasurface of $m = 4$, leading to a symmetric wave propagation behavior. We also calculate the theoretical loci of wave beams based on Eq. (1), which are shown as white dashed lines in the filed counters. In both cases, the numerical field pattern shows a perfect agreement with the theoretical prediction. We further investigate the asymmetric transmission performance for various incident angles in Fig. 2(c). A transmitted energy contrast ratio [30] is defined as $\eta = 10\ lg\ (I_N/I_P)$, where $I_P$ and $I_N$ are the energy integrals along a line parallel to the elastic metasurface with distance of $2\lambda$ on the transmitted side for PI and NI, respectively. Once the incident angle is larger than the critical angle $|\theta_c| = 16°$ (the gray shaded region), the metasurface of $m = 3$ presents a strongly asymmetric transmission phenomenon with $\eta > 5$. It is noted that $\eta$ keeps a stable high value over 10 when the incident angle is larger than 25°, because waves almost totally reflect/transmit as predicted for the case of PI/NI in such a wide range of available incident angle. On the contrary, the metasurface of $m = 4$ shows invariably symmetric transmission property with $\eta \approx 0$.

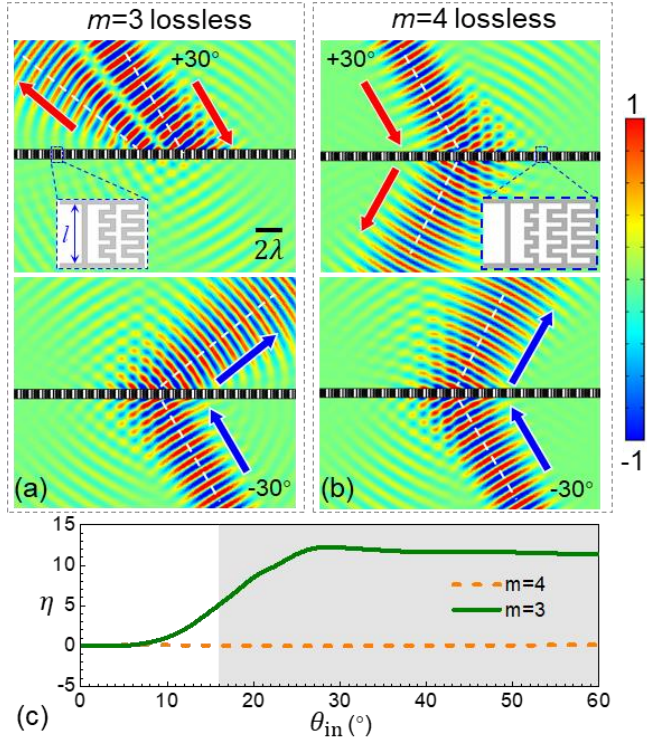

Fig. 2. Calculated transmission fields at incident angles of $\pm 30°$ through a lossless metasurface of (a) $m = 3$ and (b) $m = 4$. These two kinds of supercell designs are enlarged as insets. Theoretical loci are plotted as white dashed lines accordingly based on Eq. (1). (c) Simulated contrast ratio $\eta$ of the transmitted energy for PI and NI at different incident angles.

We further investigate the loss-induced effect of the metasurfaces on the performance of asymmetric elastic-wave transmission in Fig. 3. A loss factor of $\xi = 0.1$ on the elastic metasurface is introduced by using a complex Young's modulus $E = E_0(1 + i\xi)$. Considering that the loss-induced attenuation of high-order diffraction has been used to achieve asymmetric transmission in acoustics [30], we can expect that the structure with even unit cells ($m = 4$) will present asymmetric transmission if some loss is introduced in Fig. 3(b). The transmission is significantly attenuated due to the high-order diffraction for PI whereas the transmission for NI has only a slight reduction, exhibiting the predicted asymmetric performance. By checking its energy contrast spectrum under a series of loss factors and incident angles in Fig. 3(d), we can observe a considerable contrast ($\eta >$ 5) if the loss factor $\xi > 0.2$, but the transmitted energy will be weak under this circumstance. By contrast, the asymmetric behavior is more obvious in our proposed metasurface of $m = 3,$ as displayed of the energy density in Fig. 3(a). Compared with the case of $m = 4$, we find similar transmitted energy for NI but much weaker energy for PI here. Fig. 3(c) quantitatively reveals that the proposed metasurface ($m = 3$) gives us a higher contrast ratio for any loss factor than the contrast design of $m = 4$ in Fig. 3(d). Remarkable asymmetric elastic-wave transmission is exhibited when the incident angle is beyond $\theta_c$. The contrast ratio keeps a high value regardless of the loss factor and reaches its peak around 12 in the case of $\theta_{in} = 25°$ and $\xi = 0.2$.

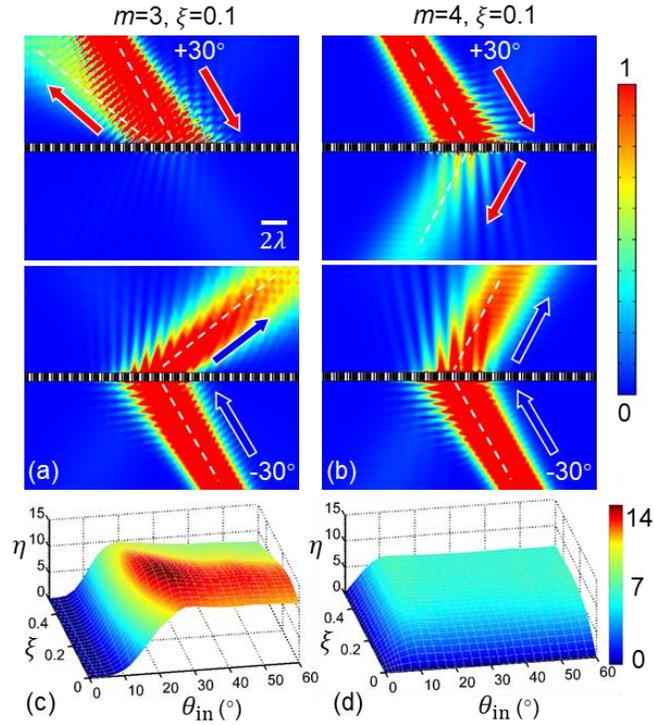

Fig. 3. (a) Calculated energy density fields through a lossy metasurface of $m = 3$ at $\theta_{in} = \pm 30°$ and $\xi = 0.1$. Theoretical loci are plotted as white dashed lines accordingly. (b) Same as (a), but for a lossy metasurface of $m = 4$. (c) Simulated energy contrast $\eta$ as a function of incident angle $\theta_{in}$ and the loss factor $\xi$ for the metasurface of $m = 3$. (d) Same as (c), but for the metasurface of $m = 4$.

The applicability and robustness of the proposed design strategy are further verified by experimental investigations. Figure 4(a) shows the experimental setup. The thin steel plate with lossless metasurface is fabricated by wire electrical discharge machining and zoomed into view as an inset of Fig. 4(a). Two arrays of 10 rectangular PZT sheets are bonded separately on plate surface with two opposite incidence angles to generate planar waves. We use a scanning laser Doppler vibrometer (Polytec, PSV-400) to capture the out-of-plane velocity field and visualize the elastic-wave propagation. A viscoelastic damping material (Blu-Tack tape) is attached to all boundaries to reduce reflections. Detailed experimental procedures are introduced in Supplementary Material [32].

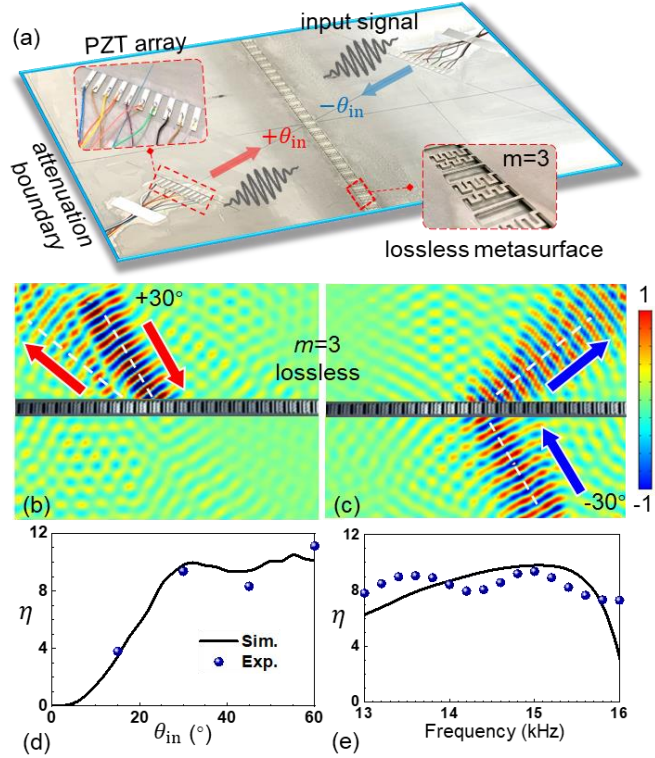

Fig. 4. (a) Photography of the fabricated sample of the lossless metasurface. Two arrays of PZT sheets are bonded separately with two opposite incident angles to generate planar waves. Experimentally measured transmission fields through the lossless metasurface of $m = 3$ (b) for PI and (c) for NI at 15 kHz. (d) Experimental and simulated energy contrasts at 15 kHz with different incident angles. (e) Energy contrasts for $\theta_{in} = \pm 30°$ at different frequencies.

Figures 4(b) and 4(c) present the experimentally measured wave fields for the metasurface of $m = 3$ at $\theta_{in} = \pm 30°$ with the excitation frequency of 15 kHz. It is observed that the incident planar waves for PI are reflected by the lossless metasurface, but for NI they are transmitted with almost no attenuation, which coincide quite well with the theoretical loci (white dashed lines). Corresponding transient responses of the whole propagation process are experimentally captured and provided in Supplementary Material [32]. It is experimentally verified that significant asymmetric elastic-wave transmission can be realized by the proposed metasurface without introducing any loss. Figures 4(d) further shows the experimentally measured and numerically calculated energy contrasts at different incident angles. To accurately describe the experimental setup, we set a 150 mm line source as the plane wave in simulations here. Very good agreement between the experimental and numerical results is observed, confirming that the proposed lossless metasurface works in a wide range of incident angles. In addition, Figure 4(e) shows the contrast ratio at different frequencies from 13-16 kHz. The value is tested quite stable around 8 within this frequency range.

To verify the advantage of the proposed oddness-induced metasurface, we tested the elastic metasurface of $m = 4$ as control experiments. Figure 5 depicts the measured energy density fields for the metasurface of $m = 4$ without and with loss at $\theta_{in} = \pm 30°$, respectively. Figure 5(a) shows that the transmission properties in the lossless metasurface of m=4 are almost symmetrical in the two opposite directions, which agrees well with the theoretical loci as plotted by dashed lines. For the first time, as shown in the transient response (see Supplementary Material [32]), we experimentally capture the delay of waves in elastic metasurface caused by multiple reflections for PI. To further validate the loss-induced attenuation by multiple reflections, a 3M acrylic foam tape (with a thickness of 0.8 mm) is pasted onto the surfaces of the whole metasurface part to introduce a tunable loss, as shown in Fig. 5(b). Without loss of generality, two/one layers of tape are pasted here onto the upper/lower surface, respectively. As demonstrated in Fig. 5(b), the loss causes a significant attenuation on the transmission for PI, but a moderate attenuation in the negative direction. As a result, asymmetric elastic-wave transmission in the loss-induced metasurface of m=4 is experimentally realized as well, but the transmitted energy is observed much weaker than the case in Fig. 4.

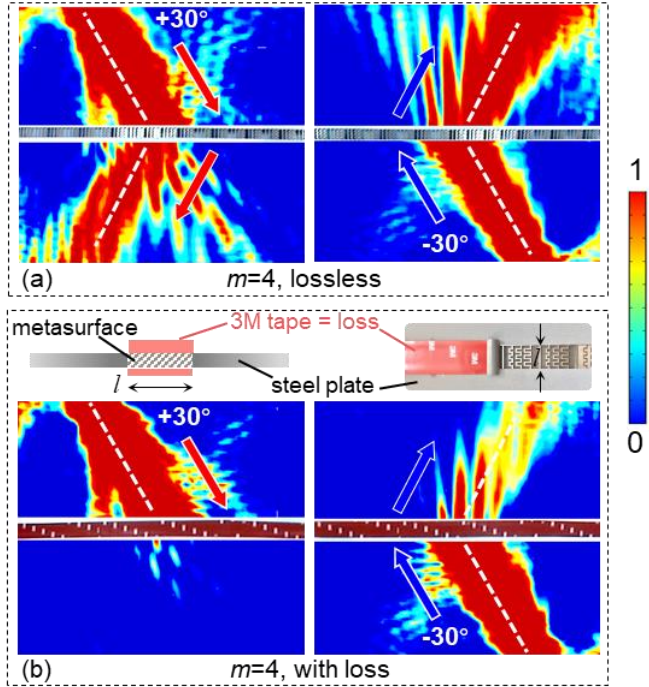

Fig. 5. Experimentally measured energy density fields at $\pm 30°$ of incidence for the metasurface of $m = 4$ (a) without loss and (b) with loss. Theoretical loci are plotted as white dashed lines accordingly. Subplots in (b) show the schematic of adding loss by attaching foam tapes to the metasurface region.

In conclusion, we have developed a generic framework for asymmetric elastic-wave transmission, within a broad range of incident angles and frequency band, through lossless oddness-induced elastic metasurfaces. The asymmetric transmission behavior can be manipulated by tailoring multiple reflections of different diffraction orders from two opposite incidences. Different from previous loss-induced metasurfaces which show an obvious attenuation of transmitted waves on both sides [30], the present design realizes nearly total reflection for PI meanwhile total transmission for NI. As a consequence, the oddness-induced metasurface numerically and experimentally exhibits a higher contrast ratio of transmission compared with the loss-induced one. We further prove that the oddness-induced metasurface still works when there is dissipation, and the contrast ratio is roughly independent of the loss factor. The proposed prototype paves the way to design compact directional devices for elastic, acoustic and other waves.

**Acknowledgements**

B.L. acknowledges financial support from the National Natural Science Foundation of China (NSFC) under grant No. 11902262. Y.L. thanks financial support from NSFC under grant No. 11902239. Z. L. acknowledges financial support from NSFC under grant No. 11672004.